\documentclass[prl,twocolumn,showpacs]{revtex4}

\usepackage{graphicx}
\usepackage{dcolumn}
\usepackage{bm}

\begin{document}


\title{Electron scattering in atomic force microscopy experiments.
       }

\author{Linda A. Zotti and Werner A. Hofer}\email{whofer@liverpool.ac.uk}
 \affiliation{
         SSRC, Department of Chemistry,
         University of Liverpool, Liverpool L69 3BX, Britain.}
\author{Franz J. Giessibl} \affiliation{
         Universit\"at Augsburg, Institute of Physics, EKM, EP 6, 86135
Augsburg, Germany.}
\begin{abstract}
It has been shown that electron transitions, as measured in a
scanning tunnelling microscope (STM), are related to chemical
interactions in a tunnelling barrier. Here, we show that the shape
and apparent height of subatomic features in an atomic force
microscopy (AFM) experiment on Si(111) depend directly on the
available electron states of the silicon surface and the silicon AFM
tip. Simulations and experiments confirm that forces and currents
show similar subatomic variations for tip-sample distances
approaching the bulk bonding length.
\end{abstract}

\maketitle

The intense research into improving the local and energy
resolution in scanning probe microscopes (SPM) over the last two
decades have made the assignment of atomic features to the
position of surface ions, at least in principle, a routine task.
The very high resolution of individual features, today below the
limit of one \AA ngstrom, makes it possible to carry out detailed
studies of the electronic structure and their local extension,
which a decade ago have been thought impossible. Of particular
importance, experimentally, was the advent of the atomic force
microscope (AFM), since AFM experiments do not rely on the
transition of electrons through the tunnelling barrier, and can
therefore be in principle be performed on all materials. Today,
AFM is able to reveal finer details than scanning tunneling
microscopy (STM) \cite{giessibl00,sim_2003,eguchi04,giessibl04}.
Theoretically, the advance in quantitative models and their
predictive power made it possible to analyze the differences
between simple theoretical models and the obtained experimental
results: by gradually obtaining a more realistic view of the
experimental situation, the role of different interactions and
their effects on SPM images was determined. The simple initial
models could thus be corrected e.g. for the effects of atomic
relaxations, of ionic charge, or dissipation processes during AFM
scans \cite{hofer03a}.
\begin{figure}[sim_cont]
\includegraphics[width=\columnwidth]{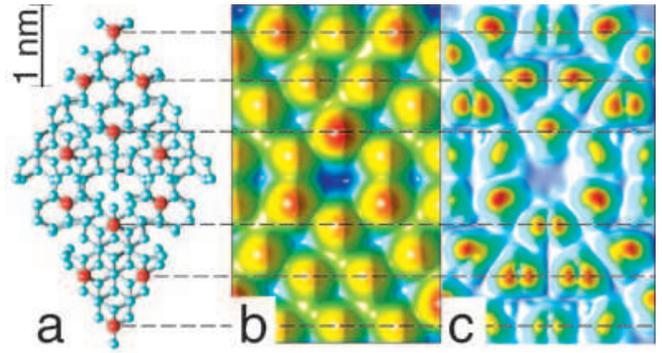}
\caption{(Color online) (a) Position of atoms in the surface layer
(blue) and adatoms (red) in the final iteration. (b) Constant
density contour of all electron states in the interval $[E_F -
2eV,E_F]$ at a median distance of 3.5\AA. Atoms are seen as single
protrusions. (c) Contributions of electron states at an energy value
of $E_F - 2$eV. Double features of states at the silicon surface in
this case are clearly visible.} \label{fig1}
\end{figure}

In this Letter we show that the subatomic features, observed by AFM
measurements on Si(111)(7$\times$7), which have been the subject of
intense debate \cite{sim_2003,hug00,lantz01}, are due to the
scattering of electrons from a very narrow energy range at the
surface into double dangling bonds of the silicon AFM-tip.

 The Si(111) surface was simulated by a four layer silicon film,
 the bottom layer was passivated with hydrogen. The ionic positions were determined
 by fully relaxing the surface layers until the Hellman-Feynman forces on individual
 ions were less than 0.01eV/\AA. The electronic groundstate was then
 calculated using ultrasoft pseudopotentials for the ionic cores
 with the Vienna ab-initio Simulation Program \cite{kresse93,kresse96}.
 Due to the size of the system, which comprises 249 ions in the unit
 cell, the calculation was limited to only one k-point, the $\bar{\Gamma}$
 point. The energy cutoff for the plane-wave expansion was set to
 250\,eV. The atomic arrangement in the final iteration and the
 (7$\times$7) unit cell is shown in Fig. \ref{fig1}(a). After the
 electronic groundstate structure had been calculated the
 Kohn-Sham states in the energy interval $E_F \pm 3$eV were
 extracted. Testing the resolution of the ensuing images based on
 the electronic surface structure alone we found that the detailed
 features are quite sensitive to the cutoff in the two dimensional
 Fourier expansion of the single electron states. To obtain high
 resolution images it was necessary to expand every state with
 close to 500 reciprocal lattice vectors. The density of electron
 charge within the interval $[E_e - 2\textrm{eV},E_F]$ in a horizontal plane
 above the surface is shown in Fig. \ref{fig1}(b). The contributions to the charge density
 at -2\,eV are shown in (c). It can be seen that the differential contributions
 show a double feature at the position of the adatoms, which is missing in the density contour over
 the whole bias interval. This indicates that the double features are due to
 electron states with an energy close to -2\,eV relative to the Fermi level. An integration
 over the whole bias range sums up a large number of individual states, the double feature,
 which we attribute to single electron states, is then no longer observable.

 The tip model in our simulations is
 a (2$\times$2) Si(001) surface with a single silicon atom, as suggested by an
 analysis of high resolution
 AFM experiments \cite{giessibl00}. The tip was simulated by an eight
 layer film with a hydrogen passivated bottom layer. Since the tip
 system is somewhat smaller we increased the number of k-points of
 the tip system to nine k-points near the center of the tip Brillouin
 zone, including the $\bar{\Gamma}$ point. The single silicon atom at the tip apex leads to two
 dangling bonds protruding from the tip apex. We simulated two STM tips,
 obtained by rotating the atoms of the original cell by 90 degrees
 while keeping the lattice vectors constant. To obtain the
 current maps and the locally resolved $dI/dV$ spectra of the
 surface we employed a newly developed method, described in
 \cite{hofer05a,palotas05}. The scattering method, calculated to
 first order in the interface Green's function accounts for the
 bias dependency with the help of an additional term, which
 depends on the bias potential $eV$ and the exponential decay of
 surface and tip states $\kappa_{i(k)}$, according to
 \cite{palotas05}:
 \begin{eqnarray}
I(V)&=&\frac{4\pi
e}{\hbar}\sum\limits_{ik}\left[f\left(\mu_S,E_k-\frac{eV}{2}\right)-f\left(\mu_T,E_i+\frac{eV}{2}\right)\right]
\nonumber\\&\times&
\left|\left(\frac{\hbar^2}{2m}-\frac{eV}{\kappa_k^2-\kappa_i^2}\right)
M_{ik}\right|^2\delta(E_i-E_k+eV).
\end{eqnarray}

Here, $f$ is the Fermi distribution function, $\mu_{S(T)}$ is the
chemical potential of the surface (tip), $E_{i(k)}$ the eigenvalue
of surface (tip) states, and $M_{ik}$ the Bardeen matrix element.
The difference to the standard Bardeen method, the second term in
the quadratic expression, arises from the shift of eigenstates due
to the applied bias voltage. It accounts for the bias dependency
in the tunnelling junction. High resolution spectra for the bias
range $\pm 2$V were directly calculated by a map of differential
changes as the bias voltage is ramped up. The method, described in
\cite{hofer05a} is more efficient than numerically differentiating
the $I(V)$ map, it also allows to separate contributions due to
the bandstructure of the STM tip.

\begin{figure}[sim_cont]
\includegraphics[width=\columnwidth]{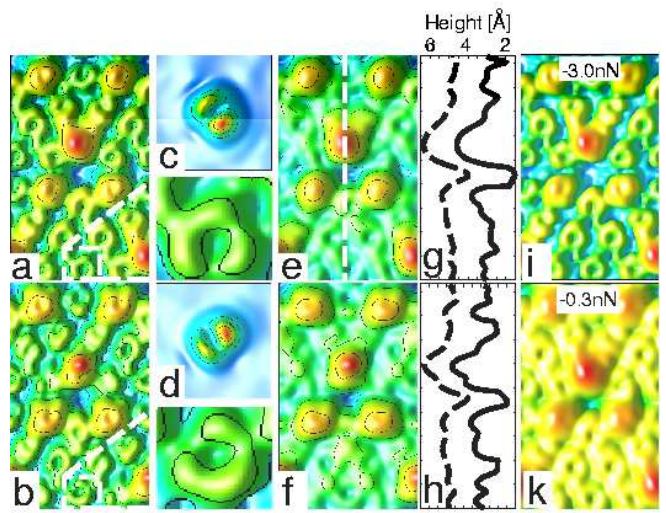}
\caption{(Color online) Constant current contour at -2\,V, simulated
for two different silicon tips in (001) orientation. (a) and (b)
Constant current contours at a current value of 0.5\,nA. The double
features at the surface are clearly visible, the rotated tip cell
leads to a rotation of features in the image (see details). (c) and
(d) Constant density contour of the two tip unit cells at plus 2eV.
The double dangling bond is rotated by 90 degree. (e) and (f)
Constant current contour at a current value of 20\,pA. The double
features at this distance are difficult to observe, and there is no
substantial difference between the images obtained with different
tips. (g) and (h) Linescans and apparent height at the center of the
cell (see dashed white line). The current values are 0.5nA (solid
lines), and 20\,pA (dashed line). (i) and (k) Constant force
contours of 3.0nN (i), and 0.3nN (k). The images are very similar to
the constant current contours.} \label{fig2}
\end{figure}

The experiments were conducted with a combined scanning
tunneling/atomic force microscope at room temperature in an
ultrahigh vacuum at $p \approx 10^{-10}$\,mbar. The Si(111) sample
(p-doped Si(111) wafer, resistivity 9.0\,$\Omega$cm, MacTecK,
J\"ulich, Germany) was prepared with a standard technique
involving heating to 1280\,$^{\circ}$C for about 30\,s. In order
to facilitate comparison between theory and experiment,
simultaneous force and current measurements are performed in the
constant-height mode, where the tip scans in a plane parallel to
the surface. Measuring the tip-sample forces directly is
impractical because of experimental noise and strong background
forces \cite{Giessibl:2003}. Therefore, we resort to frequency
modulation force microscopy \cite{albrecht91} where the frequency
shift of an oscillating cantilever is measured. We used a qPlus
cantilever\cite{Giessibl:2000b} in this study with a spring
constant of $k=1800$\,N/m, an eigenfrequency of $f_0=11886$\,Hz
and an oscillation amplitude of $A=9.4$\,\AA. While the tip of the
force sensor consisted of an etched tungsten tip, we found that
the active tip cluster consists of Si picked up from the surface
during the tip preparation process that involves controlled
collisions with the Si sample accompanied by voltage pulses. After
this process, we frequently find craters on the Si surface where
the missing sample material is not found anywhere close on the
surface. Scanning electron microscopy images of these \lq\lq
tungsten\rq\rq tips reveal sharp-edged clusters with a size
similar to the observed craters on the surface ($\approx
100$\,nm). Because the tip oscillates, the experimental observable
is an average tunneling current $I_{av}$ and a frequency shift
$\Delta f$. $I_{av}$ is given by the peak current divided by
$\sqrt{4\pi\kappa A}$ where $\kappa$ is the decay rate of the
tunneling current (Eq. 43 in \cite{Giessibl:2003}). With $\kappa
\approx 1$\,\AA$^{-1}$ and $A$ as given above, we find a peak
current of 10.8 times the average values given in the captions of
Fig. 3. The frequency shift is given by $\Delta f /f_0 k
A^{3/2}\approx 0.4 F_{ts}\sqrt{\lambda}$ where the range $\lambda$
of the tip-sample force $F_{ts}$ is given by $\lambda=1/(2\kappa)$
\cite{hofer03a}. The typical frequency shift in this experiment is
-5\,Hz, thus the estimated tip-sample force at closest distance is
-5\,nN. The results of topographic simulations at a bias voltage
of -2V are shown in Fig. \ref{fig2}.

\begin{figure}[exp_cont]
\includegraphics[width=\columnwidth]{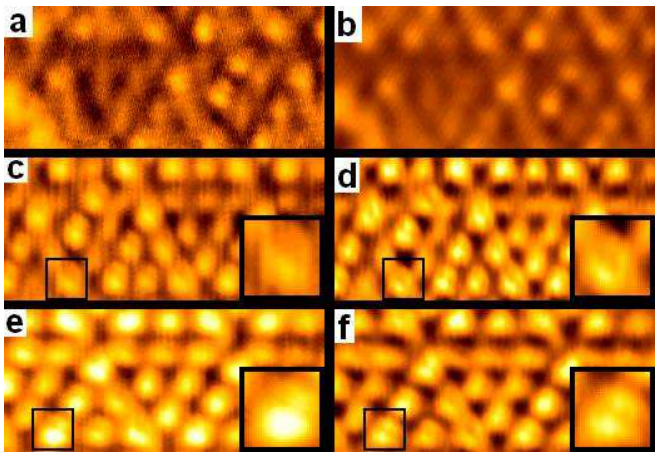} \caption{(Color
online) Simultaneous constant height images of tunneling current
(left column) and frequency shift (right column). Sample bias
-1.2\,V, average tunneling current 0.1\,nA (a), 0.5\,nA (c) and
1.0\,nA (e). The frequency shifts show a double feature at very
close distances, the orientation of the double features is similar
to the case simulated in Fig. 2 d. The insets in c-f show 2x
magnified views of one single adatom depicted by the square frames
in the left bottom regions showing subatomic details.}
\label{fig3}
\end{figure}

Simulations were performed with the tip unit cell of the extended
Si(001) tip system aligned to the horizontal and vertical axes of
the surface unit cell. For high current values of 0.5nA the
topographic images show a double feature: two protrusions with a
depression in vertical alignment for all adatoms of the surface,
with the exception of one corner atom (Fig. \ref{fig2}(a)). The
stability of the numerical results was analyzed by performing the
simulations with 100, 200, or 488 lattice vectors. The results given
are obtained with the expansion of highest resolution, but even in
this case the double feature at one corner atom, which is clearly
visible in high-resolution spectrum simulations (see further down),
could not be resolved. In all other cases, however, the constant
current contour reveals double features (Fig. \ref{fig2}(a) and
(b)). The details of the features depend on the orientation of the
tip, as a comparison between Fig. \ref{fig2}(a) and (b) reveals. The
detail shows the shape of one protrusion, which is rotated if the
tip rotates by ninety degrees. It is interesting to note that detail
(d) is actually not only rotated, but also reflected compared to
detail (c). The additional mirror plane reflects the reflected
vertical orientation of the STM tip.
For comparison, we also show constant force contours of the surface
for 3.0nN and 0.3nN (Fig. ref{fig2} (i) and (k)), and one of the tip
models.
The simulations establish that the origin of the double feature is
the convolution of surface states at the Si(111) surfaces and states
at the mono-atomic Si(001) tip. By contrast, a simulated scan of the
same surface with a tungsten tip does not show any double features
even at very close distances. This is quite understandable, if one
considers the spatial distribution of dangling bonds on the front
atom of the silicon tip, which is much more focussed than that of
the electron states at the apex of a tungsten tip. In addition, the
local resolution is diminished by a large number of nearly
equivalent tip states, which can be used as a drain for tunnelling
electrons.

\begin{figure}[sim_spec]
\includegraphics[width=\columnwidth]{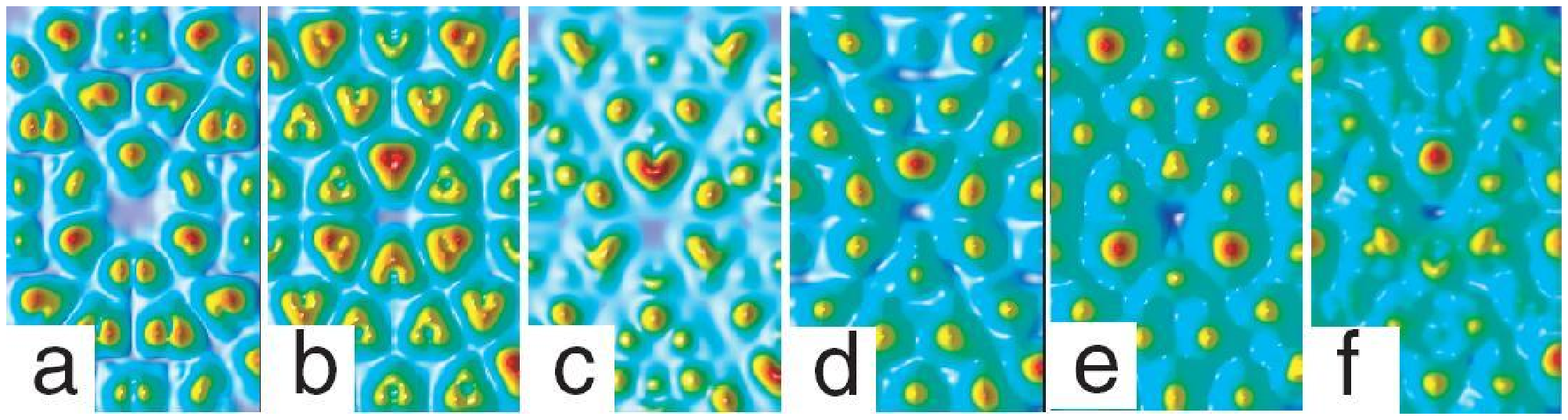}
\caption{(Color online) Conductance contours $dI/dV (x,y,z_0)$ for
negative bias voltages of -2.0V, -1.5V, and -1.0V (frames (a) to
(c)); and for positive bias voltages +1.0V, +1.5V, and +2.0V
(frames, (d) to (f)) The double features in the conductance maps
are only visible in the negative bias regime. In the positive
range atoms are imaged only as single protrusions.} \label{fig4}
\end{figure}

The occurrence of double features in the constant current contours
is distance dependent. Figs. \ref{fig2}(e) and (f) show constant
current contours of 20\,pA. At this current, which is roughly in
the range of most STM experiments, the double features at the
positions of silicon adatoms have nearly vanished. The median
distance at this current value, shown in the linescans (g) and
(h), is still only about 4\,\AA, which is a range that is too
close for conventional STM. It seems that even with a silicon tip
apex in (001) orientation, the features are difficult to observe
in conventional STM experiments. To sum up the argument, we may
say that even though the double features pertain to the electronic
structure of the silicon surface, they are only visible in dynamic
STM \cite{herz03} or dynamic AFM with a very peculiar choice for
the tip. The reason, that dynamic STM/AFM can scan at distances of
less than 3\,\AA, i.e. in a range, where the tip of conventional
STM is broken, is the near absence of shear forces on the tip. An
equivalent situation in STM experiments would be an approach and
retraction cycle, and in this case it has been shown that the STM
tip can approach to the point of contact without breaking
\cite{cross98,limot05}.

In Fig. \ref{fig3} we show simultaneously acquired images of the
current averaged over one oscillation (Fig. \ref{fig3}(a), (c), and
(e)) and the frequency shift (Fig. \ref{fig3}(b), (d), and (f)) in a
combined AFM/STM experiment. Up to a current of
0.5nA, which is well above the level in STM scans
without an oscillating cantilever,
the current contours do not show a double feature in the images,
even though the protrusions correspond to the protrusions in the
frequency shift image.

This indicates, as already pointed out in a previous paper
\cite{hembacher04}, that current and interaction energy are indeed
related. But considering that the double feature is present in the
corner atoms at very close distances (Fig. \ref{fig3}(f)), it also
reveals that the relation is not strictly linear at very low
distances. The frequency shift in this case seems more sensitive to
the details of the electronic structure than the current contour. In
addition, the different orientation of the double features points to
a rotation of the tip compared to the models used in the
simulations.

In principle, the choice of the AFM tip model should also play a
role in AFM simulations, if the resulting forces are simulated by
moving a tip cluster above the silicon surface and calculating the
ensuing interactions of the coupled system by density functional
theory \cite{sim_2003}. However, there is one qualification:
interactions and forces depend on the transition of electrons from
one side of the coupled system to the other \cite{hofer03b}. If
the energy difference between individual electron states of the
decoupled subsystems is too large, the transmission probability
will be reduced and the interaction energy will not be strongly
affected. This means that the states of surface and tip, leading
to the double features, must possess similar energies. The
conjecture was checked by spectroscopy simulations with high local
resolution, using the same silicon tip. Selected results of the
obtained $dI/dV$ maps are shown in Fig. \ref{fig4}. It can be seen
that differential contributions show the double feature primarily
in the bias range below -1.5\,V (Fig. \ref{fig4}(a) and (b)). In
the positive bias regime the adatoms are uniquely imaged as single
protrusions (Fig. \ref{fig4}(d), (e), and (f)). Comparing with
experimental conditions in most STM images of the Si(111) surfaces
it seems that this characteristic of images is one of the reasons
for choosing primarily positive bias voltages corresponding to
electrons tunnelling from filled tip states into empty surface
states. Around the Fermi level the conductance and also the
obtained current values are very low. However, we also note double
features in the conductance contours in this range. This seems to
explain to some extent the results of recent simulations. As AFM
simulations are performed with zero bias, the transitions
responsible for chemical interactions \cite{hofer03a} involve
primarily states in this range, which could lead to double
features, even though their corrugation will be significantly
smaller than in the negative bias regime. We may conclude from
these results that the experimental data, obtained by AFM
\cite{giessibl00}, which were taken at a sample bias of -1.5\,V,
utilized one range, where the double features appear. By contrast,
the simulations, performed at zero bias, correspond to a different
range. The whole picture only emerges, when the contributions to
the tunnelling currents at varying bias are analyzed in detailed
images with high local resolution. One may extend the analysis by
a conjecture concerning AFM scans on insulating surfaces. Since
the overlap between surface and tip states plays an essential role
in the obtained interaction energy, tuning the bias voltage in AFM
scans to a suitable range may substantially improve the resolution
also in these cases. Arai and Tomitori have recently found, that a
strong bias dependence also occurs in AFM imaging at greater
tip-sample distances \cite{arai04}.

In summary we have shown that the sub-atomic features arise from a
convolution of the electronic structure of the silicon surface and
a silicon tip that is highly bias dependent. These features can be
observed by \emph{both} STM and AFM, if the tip-sample distance is
sufficiently small. Stability considerations require the use of
oscillating tips to reduce lateral forces when scanning very close
to the sample. The occurrence of double peaks for single Si atom
images is also limited to two narrow bias ranges: below about
-1.5\,V, and near the Fermi level in the positive bias regime.

LAZ is funded by EPSRC grant GR/T18592/01. WAH is supported by the
Royal Society through a University Research Fellowship. The
experimental work has been funded by BMBF grant EKM13N6918. We
also acknowledge help from O. Paz and J. Soler with the setup of
the Si(111) (7$\times$7) unit cell.

\end{document}